\documentclass[fleqn,10pt]{wlscirep}
\usepackage[utf8]{inputenc}
\usepackage[T1]{fontenc}
\DeclareTextSymbol{\degre}{T1}{6}
\usepackage{siunitx}    
\usepackage{stmaryrd}   

\title{FaBiAN: A Fetal Brain magnetic resonance Acquisition Numerical phantom}

\author[1,2,*]{H\'el\`ene Lajous}
\author[1,+]{Christopher W. Roy}
\author[1,3,4,+]{Tom Hilbert}
\author[1,2]{Priscille de Dumast}
\author[1]{S\'ebastien Tourbier}
\author[1]{Yasser Alem\'an-G\'omez}
\author[1,2]{Jérôme Yerly}
\author[4]{Thomas Yu}
\author[1,2]{Hamza Kebiri}
\author[5,6]{Kelly Payette}
\author[1,2]{Jean-Baptiste Ledoux}
\author[1]{Reto Meuli}
\author[1]{Patric Hagmann}
\author[5,6]{Andras Jakab}
\author[1]{Vincent Dunet}
\author[1]{M\'eriam Koob}
\author[1,3,4,§]{Tobias Kober}
\author[1,2,§]{Matthias Stuber}
\author[2,1]{Meritxell Bach Cuadra}
\affil[1]{Department of Radiology, Lausanne University Hospital (CHUV) and University of Lausanne (UNIL), Lausanne, Switzerland}
\affil[2]{CIBM Center for Biomedical Imaging, Switzerland}
\affil[3]{Advanced Clinical Imaging Technology (ACIT), Siemens Healthcare, Lausanne, Switzerland}
\affil[4]{Signal Processing Laboratory 5 (LTS5), Ecole Polytechnique F\'ed\'erale de Lausanne (EPFL), Lausanne, Switzerland}
\affil[5]{Center for MR Research, University Children's Hospital Zurich, University of Zurich, Zurich, Switzerland}
\affil[6]{Neuroscience Center Zurich, University of Zurich, Zurich, Switzerland}

\affil[*]{helene.lajous@unil.ch}

\affil[+,§]{These authors contributed equally to this work.}


\begin{abstract}
Accurate characterization of \emph{in utero} human brain maturation is critical as it involves complex and interconnected structural and functional processes that may influence health later in life. Magnetic resonance imaging is a powerful tool to investigate equivocal neurological patterns during fetal development. However, the number of acquisitions of satisfactory quality available in this cohort of sensitive subjects remains scarce, thus hindering the validation of advanced image processing techniques. Numerical phantoms can mitigate these limitations by providing a controlled environment with a known ground truth. In this work, we present FaBiAN, an open-source Fetal Brain magnetic resonance Acquisition Numerical phantom that simulates clinical T2-weighted fast spin echo sequences of the fetal brain. This unique tool is based on a general, flexible and realistic setup that includes stochastic fetal movements, thus providing images of the fetal brain throughout maturation comparable to clinical acquisitions. We demonstrate its value to evaluate the robustness and optimize the accuracy of an algorithm for super-resolution fetal brain magnetic resonance imaging from simulated motion-corrupted 2D low-resolution series as compared to a synthetic high-resolution reference volume. We also show that the images generated can complement clinical datasets to support data-intensive deep learning methods for fetal brain tissue segmentation.
\end{abstract}
\begin{document}

\flushbottom
\maketitle
%
%
\thispagestyle{empty}


\section*{Introduction}
Today, there is a growing awareness of the importance of early brain development on health later in life\cite{volpe_brain_2009,ramirez-velez_utero_2012,bhat_autism_2014,miller_antenatal_2014,kwon_what_2017,odonnell_fetal_2017,tiemeier_closer_2017,bilder_early_2019,hayat_early_2020} as brain maturation involves complex and interconnected structural and functional processes that can be altered by various genetic and environmental factors.
Magnetic resonance imaging (MRI) may be required during pregnancy to investigate equivocal situations as a support for diagnosis and prognosis, but also for postnatal management planning\cite{gholipour_normative_2017}. In clinical routine, T2-weighted (T2w) fast spin echo (FSE) sequences are used to scan multiple 2D thick slices that provide information on the whole brain volume with a good signal-to-noise ratio (SNR) while minimizing the effects of random fetal motion during acquisition\cite{gholipour_fetal_2014}. However, stochastic movements of the fetus in the womb cause various artefacts in the images, including drops in signal intensity. Post-processing approaches built on motion estimation and correction can compensate for such artefacts. Especially, super-resolution (SR) reconstruction techniques take advantage of the redundancy between low-resolution (LR) series acquired in orthogonal orientations to reconstruct an isotropic high-resolution (HR) volume of the fetal brain with reduced intensity artefacts and motion sensitivity\cite{gholipour_robust_2010,rousseau_super-resolution_2010,kuklisova-murgasova_reconstruction_2012,kainz_fast_2015,tourbier_efficient_2015,ebner_automated_2020}. Navigating through the resulting SR volume provides valuable information on the developing brain anatomy, including consistent biometric measurements\cite{velasco-annis_normative_2015,pier_3d_2016,khawam_fetal_2021}. Besides, fetal brain tissue segmentation is critical for further investigation of brain development, especially for volumetric evaluation\cite{makropoulos_review_2018, khalili_automatic_2019,hong_fetal_2020,dou_deep_2020,payette_automatic_2021}. Manual segmentation is a cumbersome and time-consuming task. Therefore, supervised deep learning approaches that rely on annotated data have emerged as accurate techniques for automated delineation of the fetal brain\cite{khalili_automatic_2019,hong_fetal_2020,dou_deep_2020,delannoy_segsrgan_2020}. The development and validation of such advanced image processing and analysis methods require access to large-scale data to account for the subject variability, but the number of good quality, exploitable MR acquisitions available in this sensitive cohort remains relatively scarce.

Numerical simulations can mitigate these limitations by providing a controlled environment with a known ground truth for accurate, robust and reproducible research\cite{wissmann_mrxcat_2014,roy_fetal_2019}. MR developments often rely on computer simulations that enable pulse sequence design, accurate prototyping and evaluation of new advanced acquisition schemes as well as validation of reconstruction techniques in a controlled setting\cite{petersson_mri_1993,wissmann_mrxcat_2014,xanthis_high_2014,liu_fast_2017,roy_fetal_2019}. Such platforms are also valuable educational tools for physicists and technologists\cite{petersson_mri_1993,xanthis_high_2014}. In this sense, MR simulations can efficiently complement or even substitute the design and use of sophisticated experimental phantoms, as well as experiments on animal models or even on human volunteers\cite{xanthis_high_2014}.
MR simulators can be designed to address multiple challenges, such as system imperfections, multichannel transmission, correction/suppression of image artefacts, and optimization of specific absorption rate (SAR)\cite{drobnjak_development_2006,liu_fast_2017}. Motion is a major hurdle in various MRI applications, from cardiovascular MR to functional MRI analysis and fetal imaging, as it is responsible for artefacts in the images and can lead to erroneous data analysis and interpretation\cite{drobnjak_development_2006,wissmann_mrxcat_2014,xanthis_high_2014,roy_fetal_2019}. Whereas periodic movements can be directly related to physiological processes such as breathing or a heartbeat, and may therefore be compensated during post-processing, stochastic fetal motion impedes the repeatability of measurements\cite{roy_fetal_2019} and thus hinders retrospective motion correction. The difficulty of estimating such unpredictable movements results in the lack of any ground truth, yet necessary for the validation of new methods\cite{drobnjak_development_2006}. Numerical phantoms are an interesting alternative that offers a fully scalable and flexible environment where any image-acquisition, -reconstruction, or -processing technique can be evaluated, optimized, and validated from a collection of synthetic, yet realistic data that simulate multiple controlled conditions. Furthermore, the results obtained by these various strategies can be quantitatively compared to each other through simulated reference data based on full-reference image quality assessment metrics such as the mean squared error (MSE), the peak signal-to-noise ratio (PSNR), or even the more perceptual structural similarity index (SSIM)\cite{wang_image_2004,gholipour_robust_2010,sara_image_2019}.

In contrast to a simplified analytical description of the MR signal arising from proton isochromats, advanced developments in the field of MRI require more realistic numerical simulations\cite{liu_fast_2017}. Two main approaches have been investigated: on one hand, (i) the analytical numerical formalism is based on a mathematical description of both the anatomy and the MR experiment as in the Shepp-Logan head phantom\cite{shepp_fourier_1974}, and on the other hand, (ii) voxel-based phantoms are usually derived from segmented clinical acquisitions relevant to the targeted application\cite{wissmann_mrxcat_2014}. The correspondence between the image and the corresponding k-space is governed by the continuous Fourier transform in analytical numerical models which enables an accurate representation of k-space, whereas it is approximated by its discrete version in their voxel-based equivalents. Analytical numerical phantoms are powerful tools to study k-space truncation artefacts while voxel-based simulations can be used to also model physiological processes and motion that may alter the acquisitions\cite{wissmann_mrxcat_2014,roy_fetal_2019}. However, spatial and temporal resolutions depend on the original images from which voxel-based phantoms are derived, which may limit their use in reproducibility studies. Hybrid phantoms have been developed to leverage both approaches and overcome their respective limitations, resulting in versatile and realistic models as in the Mathematical Cardiac-Torso (MCAT), the extended Cardiac-Torso (XCAT) and the MRXCAT phantoms\cite{wissmann_mrxcat_2014,roy_fetal_2019}. Depending on the targeted application, advanced phantoms can be built on these models to include more features as in the case of the fetal extended Cardiac-Torso cardiovascular magnetic resonance imaging (Fetal XCMR) phantom that combines two independent XCAT models of both the anatomy and physiology of a mother and her baby with a simulation framework for 2D cardiovascular magnetic resonance (CMR) acquisitions\cite{roy_fetal_2019}. Paradoxically, the diversity of MRI simulators makes it difficult to compare one setup to another\cite{wissmann_mrxcat_2014}. Examples of advanced MRI simulation platforms based on numerical solutions of the Bloch equations include the Jülich Extensible MRI Simulator (JEMRIS)\cite{stocker_high-performance_2010} which makes it possible to simulate motion during acquisition, the MRiLab\cite{liu_fast_2017} that is built on the generalized multi-pool exchange model, a more biologically relevant tissue model, and the Magnetic Resonance Imaging SIMULator (MRISIMUL)\cite{xanthis_mrisimul_2014} which uses Graphics Processing Units (GPU) resources of a computer to enable complex large-scale analysis without any simplifications of the underlying MRI model. Indeed, it is worth noticing that realistic MRI simulations are hampered by the high computational burden associated with the 3D representation of the simulated object and the introduction of motion during acquisition\cite{xanthis_high_2014,liu_fast_2017}. Although the parallelization of calculations on computer clusters allows to speed up the simulations, such advanced infrastructures are not always available.

To our knowledge, there is no simulation framework for fetal brain MRI. In this work, we present FaBiAN, an open-source Fetal Brain magnetic resonance Acquisition Numerical phantom that simulates T2w images of the \emph{in utero} developing brain built on segmented HR anatomical images from a normative spatiotemporal MRI atlas of the fetal brain\cite{gholipour_normative_2017}. It relies on the extended phase graph (EPG) formalism\cite{hennig_calculation_2004,weigel_extended_2015,malik_extended_2018} of the signal formation, a surrogate for Bloch equations to describe the magnetization response to various MR pulse sequences, including complex acquisition schemes that involve multiple radiofrequency pulses and gradients. Modeling the evolution of spin magnetization depending on tissue properties allows to gain insight into the obtained MR signal and to evaluate its behavior. Originally suggested to assess signal intensities in multi-echo experiments with variable flip angles\cite{hennig_multiecho_1988}, the EPG algorithm has aroused growing interest in recent years, especially for precise characterization of echoes\cite{hennig_hyperechoes_2001,hennig_multiecho_2003,weigel_contrast_2006} and diffusion effects\cite{weigel_extended_2010}, evaluation of physics-constrained reconstruction methods\cite{tamir_computational_2020}, sequence pulse design\cite{hennig_calculation_2004,sbrizzi_optimal_2016,keerthivasan_efficient_2019} and quantitative MRI techniques\cite{prasloski_applications_2012,lankford_fast_2015,cloos_multiparametric_2016}. The EPG formalism relies on the Fourier representation of the evolution of spin magnetization within a voxel after application of various RF pulses and gradients. It assumes that a single set of relaxation parameters can characterize a given tissue\cite{malik_extended_2018}. The Fourier series are used to account for the multiple resonant frequencies that may arise from local magnetic field inhomogeneities within a voxel. The EPG algorithm provides fast and accurate simulations where the resolution of Bloch equations may be demanding due to the need to compute a numerical solution for each resonant frequency within a single voxel.

Simulations of FSE acquisitions of the fetal brain by FaBiAN are based on a flexible and realistic setup that accounts for intensity non-uniformity fields and stochastic fetal motion. We investigate the capabilities of the developed framework and provide a proof of concept of its practical value in two key application examples. First, we generate LR images with multiple levels of motion to fine-tune a SR reconstruction algorithm\cite{tourbier_efficient_2015,tourbier_medical-image-analysis-laboratorymialsuperresolutiontoolkit_2020}. Then, we explore the potential of using multiple fetal brain images simulated at different gestational ages (GA) with several motion amplitudes and a variety of acquisition parameters for data augmentation in fetal brain tissue segmentation.

\section*{Methods}

\subsection*{Numerical implementation of FSE sequences}

\begin{figure}[hbt!]
\centering
\includegraphics[width=0.8\linewidth]{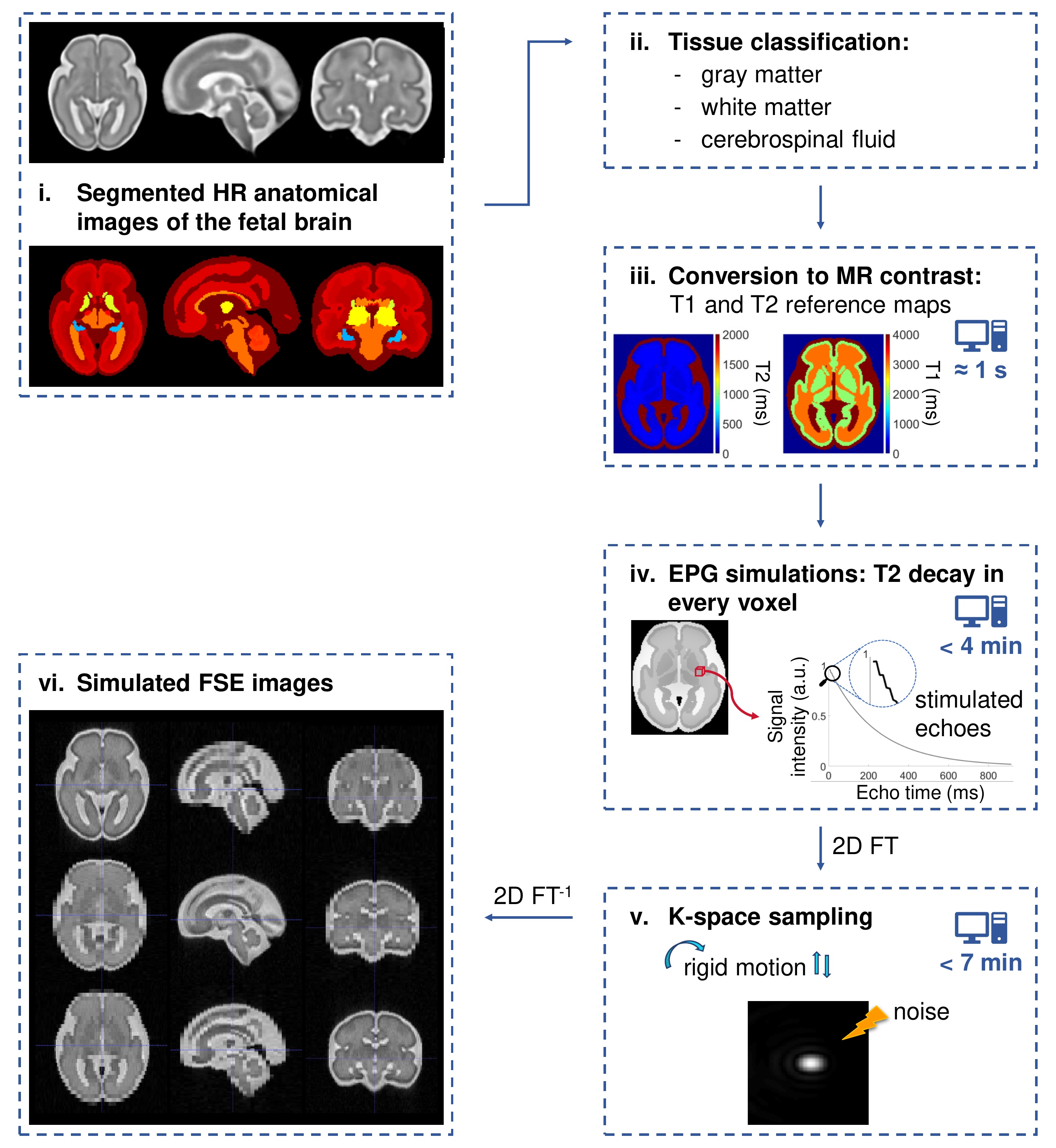}
\caption{Workflow for simulating images of the fetal brain acquired by a FSE sequence (i) from segmented HR anatomical MR images\cite{gholipour_normative_2017}, illustrated for a fetus of 30 weeks of GA. (ii) Brain tissues are organized into gray matter, white matter and cerebrospinal fluid. (iii) Anatomical structures are converted to the corresponding MR contrast to obtain reference T1 and T2 maps of the fetal brain at either 1.5 or 3 T. (iv) The EPG algorithm\cite{weigel_extended_2015,weigel_epg_2021} allows to accurately simulate the T2 decay over time in every brain voxel by accounting for the effects of the stimulated echoes, as highlighted by the enlargement of the beginning of the curve. This spatiotemporal information is subsequently used (v) to sample the Fourier domain of the simulated images of the moving fetus. After the addition of noise to match the SNR of real clinical acquisitions, (vi) synthetic FSE images of the fetal brain are eventually recovered by 2D inverse Fourier transform ($2D$ $FT^{-1}$).}
\label{fig:simulation_pipeline}
\end{figure}

Figure \ref{fig:simulation_pipeline} provides an overview of the workflow implemented in MATLAB (MathWorks, R2019a) to simulate T2w fetal brain images acquired using clinical FSE sequences. i) High-resolution anatomical images from a normative spatiotemporal MRI atlas\cite{gholipour_normative_2017} are used as a model of normal fetal brain. ii) Segmented brain tissues are organized into gray matter, white matter and cerebrospinal fluid as shown in Table \ref{tab:tissue_classification}, and iii) are assigned relaxometry properties from the literature at 1.5 T\cite{hagmann_t2_2009,nossin-manor_quantitative_2013,blazejewska_3d_2017,yarnykh_quantitative_2018,vasylechko_t2_2015} or 3 T\cite{stanisz_t1_2005,rooney_magnetic_2007,shin_fast_2009,bojorquez_what_2017,daoust_transverse_2017} accordingly. iv) The T2 decay over time is computed in every voxel of the HR anatomical images from the sequence parameters using the EPG formalism\cite{weigel_extended_2015,weigel_epg_2021}. v) The Fourier domain, or k-space, of the simulated images is sampled from the T2 decay matrix to reflect the process of FSE acquisition in the presence of random rigid motion. vi) The final simulated images are recovered by a 2D inverse Fourier transform.
As highlighted by Table \ref{tab:simulation_settings}, we have developed this open-source numerical phantom with the idea of keeping the framework as general as possible to enable users a large flexibility in the type of images simulated. As such, multiple acquisition parameters can be set up with respect to the MR contrast (effective echo time, excitation/refocusing pulse flip angles, echo spacing, echo train length), the geometry (number of 2D slices, slice orientation, slice thickness, slice gap, slice position, phase oversampling), the resolution (field-of-view, matrix size), the resort to any acceleration technique (acceleration factor, number of reference lines) or to scanner interpolation, as well as other settings related to the age of the fetus, the radiofrequency transmit field inhomogeneities, the amplitude of random fetal motion in the three main directions, and the SNR. The entire simulation pipeline is described in detail in the following.
\\
\textbf{Code availability -}
FaBiAN source code is distributed under the open source BSD 3-Clause License to support its use for research purposes. It can be downloaded from the following Zenodo repository\cite{lajous_fabian_2021}.

\begin{table}[hbt!]
\centering
\begin{tabular}{l l l}
\hline
Gray matter & White matter & Cerebrospinal fluid \\
\hline
Amygdala & Cerebellum & Cerebrospinal fluid \\
Caudate & Corpus callosum & Lateral ventricle \\
Cortical plate & Fornix & \\
Hippocampus & Hippocampal commissure & \\
Putamen & Intermediate zone & \\
Subthalamic nuclei & Internal capsule & \\
Thalamus & Midbrain & \\
 & Miscellaneous & \\
 & Subplate & \\
 & Ventricular zone & \\
\hline
\end{tabular}
\caption{\label{tab:tissue_classification}Classification of segmented brain tissues\cite{gholipour_normative_2017} as gray matter, white matter and cerebrospinal fluid.}
\end{table}

\begin{table}[hbt!]
\centering
\begin{tabular}{l l}
\hline
GA (weeks) & 30 (from 21 to 38) \\
Magnetic field strength (T) & 1.5 \\
Field inhomogeneities & from BrainWeb\cite{cocosco_brainweb_1997,noauthor_brainweb_nodate} \\
Acquisition parameters \\
\hspace{+4mm} Contrast \\
\hspace{+8mm} Effective echo time ($ms$) & 90 \\
\hspace{+8mm} Echo spacing ($ms$) & 4.08 \\
\hspace{+8mm} Echo train length & 224 \\
\hspace{+8mm} Excitation flip angle (\degre{}) & 90 \\
\hspace{+8mm} Refocusing pulse flip angle (\degre{}) & 180 \\
\hspace{+4mm} Geometry \\
\hspace{+8mm} Slice orientation & sagittal, coronal or transverse \\
\hspace{+8mm} Slice thickness ($mm$) & 3.00 \\
\hspace{+8mm} Slice gap ($mm$) & 0.3 \\
\hspace{+8mm} Number of slices & 46 \\
\hspace{+8mm} Phase oversampling ($\%$) & 80.3571 \\
\hspace{+8mm} Shift of the field-of-view ($mm$) & $\pm1.6$ \\
\hspace{+4mm} Resolution \\
\hspace{+8mm} Field-of-view ($mm^2$) & $360 \times 360$ \\
\hspace{+8mm} Base resolution ($voxels$) & 320 \\
\hspace{+8mm} Phase resolution ($\%$) & 70 \\
\hspace{+8mm} Reconstruction matrix & 320 x 404 \\
\hspace{+8mm} Zero-interpolation filling & None \\
\hspace{+4mm} Acceleration technique \\
\hspace{+8mm} Reference lines & 42 \\
\hspace{+8mm} Acceleration factor & 2 \\
Amplitude of 3D rigid motion \\
\hspace{+4mm} Little motion \\
\hspace{+8mm} Translation ($mm$) in x & $\pm1$ \\
\hspace{+8mm} Translation ($mm$) in y & $\pm1$ \\
\hspace{+8mm} Translation ($mm$) in z & $\pm1$ \\
\hspace{+8mm} 3D rotation (\degre{}) & $\pm2$ \\
\hspace{+4mm} Moderate motion \\
\hspace{+8mm} Translation ($mm$) in x & $\pm3$ \\
\hspace{+8mm} Translation ($mm$) in y & $\pm3$ \\
\hspace{+8mm} Translation ($mm$) in z & $\pm3$ \\
\hspace{+8mm} 3D rotation (\degre{}) & $\pm5$ \\
\hspace{+4mm} Strong motion \\
\hspace{+8mm} Translation ($mm$) in x & $\pm4$ \\
\hspace{+8mm} Translation ($mm$) in y & $\pm4$ \\
\hspace{+8mm} Translation ($mm$) in z & $\pm4$ \\
\hspace{+8mm} 3D rotation (\degre{}) & $\pm8$ \\
Noise \\
\hspace{+4mm} Mean & 0 \\
\hspace{+4mm} Standard deviation & 0.15 \\
\hline
\end{tabular}
\caption{\label{tab:simulation_settings}The flexibility of FaBiAN is illustrated by the number of sequence parameters and settings available to the user. Default values are given for a fetus of 30 weeks of GA and for the clinical HASTE protocol dedicated to fetal brain examination at CHUV (see Datasets).}
\end{table}

\subsubsection*{Fetal brain model and MR properties}
Our numerical phantom is based on segmented 0.8-mm-isotropic anatomical images (Fig.\ref{fig:simulation_pipeline}-i) from the normative spatiotemporal MRI atlas of the developing brain built by Gholipour and colleagues from normal fetuses scanned between 19 and 39 weeks of gestation\cite{gholipour_normative_2017}. Due to the lack for ground truth relaxometry measurements in the fetal brain, all thirty-four segmented tissues are merged into three classes according to medical experts: gray matter, white matter and cerebrospinal fluid (Fig.\ref{fig:simulation_pipeline}-ii and Table \ref{tab:tissue_classification}). Corresponding T1 and T2 relaxation times at 1.5 T\cite{hagmann_t2_2009,nossin-manor_quantitative_2013,blazejewska_3d_2017,yarnykh_quantitative_2018,vasylechko_t2_2015} are assigned to these anatomical structures to obtain reference T1 and T2 maps, respectively (Fig.\ref{fig:simulation_pipeline}-iii).
The value of these relaxometry properties at 3 T is estimated from the literature\cite{stanisz_t1_2005,rooney_magnetic_2007,shin_fast_2009,bojorquez_what_2017,daoust_transverse_2017}, assuming that doubling the magnetic field strength from 1.5 T to 3 T increases the T1 relaxation time by approximately 25\% in gray matter and 10\% in both white matter and cerebrospinal fluid, while the T2 properties remain unchanged.

\subsubsection*{Intensity non-uniformity (INU) fields}
Non-linear slowly-varying INU fields due to transmit field inhomogeneities (B1\textsuperscript{+}) are based on BrainWeb estimations from real scans to simulate T2w images\cite{noauthor_brainweb_nodate,kwan_extensible_1996,cocosco_brainweb_1997,collins_design_1998,kwan_mri_1999}. The available 20\% INU version is resized to fit the dimensions of the atlas images and normalized by 1.2 to provide multiplicative fields in the range of 0.8 to 1.2 over the brain area. It is subsampled to a 0.1-mm resolution with linear interpolation in the slice thickness orientation in order to account for B1 bias field variations across the slice profile.

\subsubsection*{EPG formalism}
The EPG algorithm\cite{weigel_extended_2015,weigel_epg_2021} simulates the T2 decay in every voxel of the anatomical images over each echo train (Fig.\ref{fig:simulation_pipeline}-iv) based on the FSE sequence pulse design. The T1 and T2 maps of the fetal brain are fed into the EPG estimation, as well as the realistic INU fields described above that results in a spatially varying flip angle. The resulting 4D matrix that combines information about both the anatomy and the magnetic relaxation properties of the fetal brain is hereafter referred to as the T2 decay matrix.

\subsubsection*{K-space sampling and image formation}
The T2 decay matrix is Fourier-transformed and subsequently used for k-space sampling of the simulated images. For a given echo time (TE), at most one line from the associated Fourier domain of the T2 decay matrix is used, with the central line corresponding to the effective TE. If resorting to any acceleration technique such as GRAPPA interpolation, multiple reference lines are consecutively sampled around the center of k-space. Beyond, one line out of two is actually needed to simulate an acceleration factor of two. As a first approximation, these sampled lines are copied to substitute the missing lines.
According to partial Fourier imaging techniques, the properties of Hermitian symmetry in the frequency domain are used to fill the entire k-space.
While intra-slice motion can be neglected in FSE sequences, inter-slice random 3D translation and rotation of the fetal brain are implemented during k-space sampling (Fig. 1-v).
If needed, zero-interpolation filling (ZIP) is implemented by filling the edges of the simulated k-space with zeros to reach the desired reconstruction matrix size. Data in k-space are previously processed using a Fermi low-pass filter with a radius of 0.85 and a width of 1/23 to avoid Gibbs ringing artefacts\cite{lowe_spatially_1997,friedman_reducing_2006}.
Complex Gaussian noise (mean, 0; standard deviation, 0.15 for Half-Fourier Acquisition Single-shot Turbo spin Echo (HASTE) implementation, 0.01 for Single-Shot Fast Spin Echo (SS-FSE) implementation respectively) is added to simulate thermal noise generated during the acquisition process and qualitatively match the SNR of clinical data.
The simulated images are eventually recovered by 2D inverse Fourier transform (Fig.\ref{fig:simulation_pipeline}-vi).

With the aim of replicating the clinical protocol for fetal brain MRI, FSE acquisitions are simulated in the three orthogonal orientations. Besides, the position of the field-of-view is slightly shifted by $\pm1.6 mm$ in the slice thickness orientation to produce additional partially-overlapping datasets in each orientation.

\subsubsection*{Fetal motion}
The amplitude of typical fetal movements is estimated from clinical data\cite{oubel_reconstruction_2012}. Three levels are defined accordingly for little, moderate and strong motion of the fetus, with a maximum of $5\%$ corrupted slices over the fetal brain volume. They are characterized by a uniform distribution of respectively $[-1,1]mm$, $[-3,3]mm$ and $[-4,4]mm$ for independent translation in every direction and $[-2,2]$\degre{}, $[-5,5]$\degre{} and $[-8,8]$\degre{} for 3D rotation (Fig.\ref{fig:simulation_pipeline}-v).

\subsubsection*{Computational specifications}
Since the addition of 3D motion during k-space sampling is expensive in computing memory, the simulations are run on 16 CPU workers in parallel with 20 GB of RAM each.

\subsection*{Simulating data from clinical MR acquisitions}

\subsubsection*{Clinical datasets}
Data were acquired in accordance with the relevant guidelines and regulations, under the supervision of Ethics Boards composed of representatives at different levels (hospitals, cantons, and federal state). Mothers of all fetuses included in the current work provided written informed consent for the re-use of their data for research purposes.
Clinical cases are used as typical acquisition examples to generate realistic synthetic images of the fetal brain throughout development and to visually compare the quality of the simulated images.

Thirteen healthy subjects in the GA range of 21 to 33 weeks were scanned at Lausanne University Hospital (CHUV) as part of a larger institutional research protocol approved by the ethical committee of the
Canton of Vaud, Switzerland (CER-VD, decision number: 2021-00124). In particular, these clinical cases are used to showcase the implemented pipeline and the realistic appearance of the corresponding synthetic HASTE images, as well as to explore the potential of FaBiAN in optimizing SR fetal brain MRI.

Fifteen subjects (thirteen neurotypical subjects and two subjects with light ventriculomegaly) from the Fetal Tissue Annotation Dataset (FeTA)\cite{payette_automatic_2021} in the GA range of 21 to 34.6 weeks were scanned at University Children’s Hospital Zurich (Kispi). Their inclusion in research studies was approved by the ethical committee of the
Canton of Zurich, Switzerland (KEK, decision number: 2016-01019).
The original LR series were reconstructed using a simplified version of the Image Registration Toolkit (SIMPLE IRTK)\cite{kuklisova-murgasova_reconstruction_2012} under Licence from Ixico Ltd, including rigid registration to template\cite{studholme_overlap_1999}. Preprocessing steps included the manual creation of a mask for the reference image series, denoising using the Baby Brain Toolkit module\cite{rousseau_btk:_2013} and bias field correction by 3D Slicer\cite{tustison_n4itk_2010,fedorov_3d_2012}.
The resulting SR reconstructions were manually segmented according to the FeTA annotation guidelines\cite{payette_automatic_2021}.
In particular, the clinical cases from Kispi allow to extend FaBiAN to generate multiple SS-FSE images of the fetal brain with various settings. We investigate if these simulated images are realistic enough to replace part of the original data in the training phase of a deep learning network for fetal brain tissue segmentation, and to complement a clinical dataset to improve the performance of the segmentation algorithm.

\subsubsection*{Clinical MRI protocol}
Typical fetal brain HASTE images are acquired on patients at 1.5 T (MAGNETOM Aera, Siemens Healthcare, Erlangen, Germany) with an 18-channel body coil and a 32-channel spine coil at CHUV. At least three T2w series of 2D thick slices are acquired in three orthogonal orientations using an ultra-fast multi-slice HASTE sequence (TR/TE, $1200 ms$/$90 ms$; flip angle, $90$\degre{}; echo train length, $224$; echo spacing, $4.08 ms$; field-of-view, $360 \times 360 mm^2$; voxel size, $1.13 \times 1.13 \times 3.00 mm^3$; inter-slice gap, $10\%$)\cite{khawam_fetal_2021,lajous_dataset_2020}. Twenty-two to thirty slices are needed to cover the whole fetal brain depending on the GA (between 21 and 33 weeks) and size of the fetus, which corresponds to an acquisition time of between 26 to 36 seconds.

Fetal brain SS-FSE images are acquired on patients on either a 1.5-T or 3-T clinical GE whole-body scanner (Signa Discovery MR450 or MR750), either using an 8-channel cardiac coil or body coil, at Kispi. At least three series of 2D thick slices are acquired in three orthogonal orientations using a T2w SS-FSE sequence (TR/TE, $3000-3200 ms$/$116.032-124.08 ms$; flip angle, $90$\degre{}; echo train length, $224$; echo spacing, $3-3.2 ms$; field-of-view, from $240 \times 240 mm^2$ to $300 \times 300 mm^2$; isotropic in-plane resolution, from $0.47 \times 0.47 mm^2$ to $0.59 \times 0.59 mm^2$; slice thickness, $3.00-4.00 mm$)\cite{payette_automatic_2021}. Twenty-four to forty-three slices are needed to cover the whole fetal brain depending on the GA (between 21 and 34.6 weeks) and size of the fetus, which corresponds to an acquisition time of between one to two minutes.

The position of the field-of-view is slightly shifted in the slice thickness orientation to acquire additional data with some redundancy. In clinical practice, a total of six partially-overlapping LR series are commonly acquired in the three orthogonal orientations for subsequent SR reconstruction of the fetal brain.

\subsubsection*{Simulated datasets}
The general framework presented in this paper makes it possible to simulate the clinical acquisition schemes described above for different MR vendors, at various magnetic field strengths, and with realistic SNR and amplitude of fetal movements.

The clinical cases from CHUV are used as representative examples of fetal brain HASTE acquisitions: the corresponding sequence parameters are replicated to simulate HASTE images of the fetal brain at various GA. The amplitude of fetal movements in clinical acquisitions is assessed by an engineer expert in MR image analysis to ensure a similar level of motion in the simulated images.
Besides, a 3D HR 1.1-mm-isotropic HASTE image of the fetal brain is simulated without noise or motion to serve as a reference for the quantitative evaluation of SR reconstructions from simulated LR 1.1-mm-in-plane HASTE images.

SS-FSE images of the fetal brain are simulated for fifteen subjects in the GA range of 21 to 35 weeks at either 1.5 T or 3 T. We reproduce the same acquisition parameters and geometry as in the clinical dataset from Kispi. For every subject, three partially-overlapping series are simulated in each of the three orthogonal orientations, two with little motion and one with moderate motion.

\begin{figure}[hbt!]
\centering
\includegraphics[width=\textwidth]{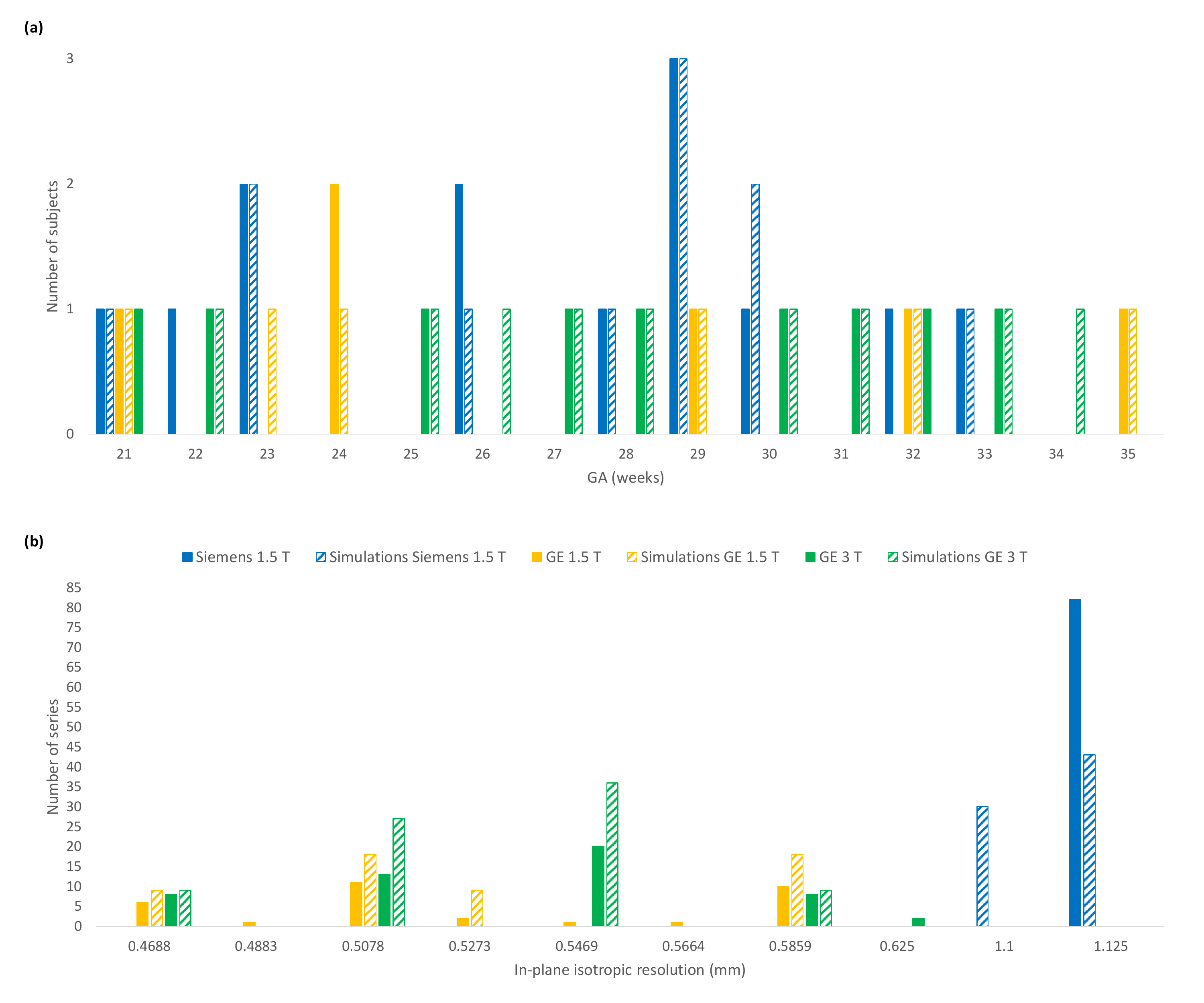}
\caption{Distribution of (a) GA across all clinical cases and simulated subjects involved in this work and (b) in-plane isotropic resolution in the corresponding LR FSE images of the fetal brain.}
\label{fig:histograms}
\end{figure}

Figure \ref{fig:histograms} displays the histograms of the distribution of GA across the original clinical cases and the simulated subjects according to the MR vendor and the main magnetic field strength, and of the in-plane isotropic resolution in the corresponding LR images of the fetal brain.

\subsubsection*{Motion index}
A motion index is defined to support the assessment of the level of fetal movements in acquisitions and simulations based on binary masks that are drawn on each LR series to cover the whole fetal brain volume\cite{yushkevich_user-guided_2006}. This index is estimated from tracking the slice-to-slice displacement of the 2D brain mask centroids. It is computed as the sum of distances between the centroids of the 2D brain masks of adjacent slices over the central third of the slices of the 3D brain mask normalized by the number of slices considered. This index was correlated with the assessment of the motion level in multiple clinical acquisitions by an engineer expert in MR image analysis. An index less than 0.5 mm stands for little motion, in the range of [0.5,1]mm for moderate motion, and larger than 1 mm for strong motion. This motion index correlates well with the assessment of motion by an engineer expert in MR image analysis and is used in the following to estimate the amplitude of fetal movements in clinical acquisitions and ensure a similar level of motion in the simulated images.

\subsection*{Qualitative assessment}

Two medical doctors specialized in neuroradiology and pediatric (neuro)radiology respectively, provided qualitative assessment of the HASTE images of the fetal brain simulated in the GA range of 21 to 33 weeks, in the three orthogonal orientations and with various levels of motion. Special attention was paid to the MR contrast between brain tissues, to the SNR, to the delineation and sharpness of the structures of diagnostic interest that are analyzed in clinical routine, as well as to characteristic motion artefacts.
To facilitate visualization and qualitative comparison of images throughout this manuscript, simulated brain images are co-registered with clinical acquisitions of fetuses at the same GA and with equivalent level of motion, using the landmark registration of 3D Slicer\cite{johnson_brainsfit_2007}.

\subsection*{Application 1: SR reconstruction}

\subsubsection*{Implementation of SR reconstruction}
Orthogonal T2w LR HASTE series contain redundant information that enables the subsequent reconstruction of a 3D HR volume.
Clinical acquisitions, respectively simulated images, are combined into a motion-free 3D image $\mathbf{\hat{X}}$ using the Total Variation (TV) SR reconstruction algorithm\cite{tourbier_efficient_2015,tourbier_medical-image-analysis-laboratorymialsuperresolutiontoolkit_2020} which solves:
\begin{equation}
\begin{aligned}
\mathbf{\hat{X}} = \arg\min_{\mathbf{X}} \ \frac{\lambda}{2} \sum_{kl} \| \underbrace{\mathbf{D}_{kl}\mathbf{B}_{kl}\mathbf{M}_{kl}}_{\mathbf{H}_{kl}} \mathbf{X} - \mathbf{X}_{kl}^{LR}\|^2 + \|\mathbf{X}\|_{TV},
\end{aligned}
\label{eq:sr}
\end{equation}
where the first term relates to data fidelity with $k$ being the $k$-th LR series $\mathbf{X}^{LR}$ and $l$ the $l$-th slice, $\|\mathbf{X}\|_{TV}$ is a TV prior introduced to regularize the solution, and $\lambda$ balances the trade-off between data fidelity and regularization terms (default setting, $\lambda$=0.75). $\mathbf{D}$ and $\mathbf{B}$ are linear downsampling and Gaussian blurring operators given by the acquisition characteristics. $\mathbf{M}$ encodes the rigid motion of slices.

Clinical acquisitions from CHUV and the corresponding simulated images are reconstructed using the docker version of the MIAL Super-Resolution Toolkit reconstruction pipeline\cite{tourbier_efficient_2015,tourbier_medical-image-analysis-laboratorymialsuperresolutiontoolkit_2020}.

\subsubsection*{Regularization setting}
LR HASTE images of the fetal brain are simulated to mimic clinical MR acquisitions of three subjects of 26, 30 and 33 weeks of GA respectively, with particular attention to ensuring that the motion level is respected. For each subject, a SR volume of the fetal brain is reconstructed from the various orthogonal acquisitions, either real or simulated (six, respectively seven and eight series were acquired and are further simulated in the subject of 26, respectively 30 and 33 weeks of GA), with different values of $\lambda$ (0.1, 0.3, 0.75, 1.5, 3) to study the potential of FaBiAN in optimizing the quality of the SR reconstruction in a clinical setup. Indeed, the weight $\lambda$ is a sensitive hyper-parameter in SR methods that are based on solving an inverse problem. A quantitative analysis is conducted on the resulting SR reconstructions to determine the value of $\lambda$ that provides the sharpest reconstruction of the fetal brain with high SNR, namely the smallest normalized root mean squared error (NRMSE) with respect to the corresponding synthetic 3D HR ground truth.

\subsubsection*{Number of LR series: an SNR and motion case study}
Static LR HASTE images of the fetal brain (GA of 30 weeks) are recovered after the addition of various levels of complex Gaussian noise (mean, 0; standard deviation, 0.07, 0.15 or 0.3) to k-space. A standard deviation of 0.15 results in synthetic images that closely resemble clinical acquisitions.
Six independent realizations of the 2D series are generated for each SNR. A SR volume of the fetal brain is reconstructed from three, six and nine orthogonal LR HASTE series using the different realizations. In total, fifty-four SR volumes are reconstructed.

Additional LR HASTE acquisitions of the fetal brain (GA of 30 weeks, standard deviation of the noise of 0.15) are simulated with inter-slice motion. The impact of little and moderate movements of the fetus on SR reconstructions from various numbers of LR series is studied using a reference series without motion and with little amplitude of fetal motion respectively.

The different configurations are compared to each other by computing the NRMSE, the local SSIM and its mean (MSSIM) over the image with respect to the 3D isotropic ground truth. The latter are computed using Matlab \emph{ssim} function with the standard deviation of the isotropic Gaussian function that determines the weights of the pixels in a neighborhood to estimate local statistics set to its default value (1.5), and the dynamic range of the normalized input images to 255.


\subsection*{Application 2: Data augmentation for automated fetal brain tissue segmentation}

LR SS-FSE images of the fetal brain are simulated based on the MR sequence parameters and acquisition settings extracted from the clinical cases scanned at Kispi to mimic typical MR acquisitions, thus minimizing confounding factors.
The synthetic images are interpolated to $0.8594mm \times 0.8594mm$ in the in-plane direction to match the resolution of the clinical SR reconstructions. Label maps are automatically generated from the simulation framework using a nearest-neighbour interpolation. Two different labels are assigned to the ventricular system (lateral, third and fourth ventricles) and to the extra-axial cerebrospinal fluid spaces in the SR reconstructions from Kispi subjects\cite{payette_automatic_2021}, whereas the segmented HR anatomical images\cite{gholipour_normative_2017} from which our simulations are derived distinguish the lateral ventricles from the cerebrospinal fluid\cite{gholipour_normative_2017}. For consistency between brain annotations, we merge the labels that correspond to structures from the ventricular system and the cerebrospinal fluid spaces in both models.

\subsubsection*{Network architecture}
We designed a convolutional neural network based on the well established U-Net architecture\cite{ronneberger_u-net_2015} for biomedical semantic image segmentation, as it recently proved its ability to perform well for 2D fetal brain MRI tissue segmentation\cite{khalili_automatic_2019}. 
The baseline 2D U-Net is trained using an hybrid loss function $\mathcal{L}_{hybrid}$.
\begin{equation}
\begin{aligned}
\mathcal{L}_{hybrid} = (1 - \lambda)\mathcal{L}_{cce} + \lambda\mathcal{L}_{dice}
\end{aligned}
\label{eq:U-Net}
\end{equation}
where \begin{math}\mathcal{L}_{cce}\end{math} is the categorical cross-entropy loss function, \begin{math}\mathcal{L}_{dice}\end{math} the Dice loss layer that maximizes the Dice coefficient between network output and manual segmentation\cite{khalili_automatic_2019,milletari_v-net_2016} to mitigate any imbalance in the samples from the different classes. \begin{math}\lambda\end{math} balances the contribution of the two terms of the loss.

Our 2D U-Net model consists of an encoding and a decoding paths. Each layer in the encoding path is composed of a cascade of double 3x3 convolutions, followed by a rectified linear unit (ReLu) activation function and a 2x2 max-pooling downsampling layer. The number of feature maps is doubled after each layer, from 32 to 512. In the decoding path, 2x2 upsampled encoded features concatenated with the corresponding encoding path are 3x3 convolved and passed through the ReLu activation function. All convolutional layer outputs are batch normalized. The network prediction is computed with a final 1x1 convolution.
The implementation is performed in the framework of TensorFlow 2.5\cite{abadi_tensorflow_2016} and an Nvidia GeForce RTX 2080 GPU is deployed for training.

\subsubsection*{Experimental design}
As indicated in Table \ref{tab:design}, we compare four configurations that combine different proportions of subjects with SR volumes from clinical MR acquisitions and subjects with simulated LR images: a baseline (\emph{A}) that consists of real data only (fifteen subjects), a configuration (\emph{B}) that gathers ten original subjects and five simulated ones, a configuration (\emph{C}) with eight original subjects and seven simulated ones, and a configuration (\emph{D}) that complements the fifteen original subjects from the baseline (\emph{A}) with fifteen simulated ones.
Since we could not automatically propagate the annotations on the simulated images to the corresponding SR reconstruction, and conversely the labels in the SR reconstruction from clinical acquisitions to the original LR 2D series, three LR series are needed to replace one SR volume. The segmentation is computed on skull-stripped images to process only the voxels within the intracranial volume.

\begin{table}[hbt!]
\centering
\begin{tabular}{c c c c}
\hline
Configuration & Number of clinical subjects & Number of simulated subjects & Total number of subjects \\
\hline
(\emph{A}) & 15 & 0 & 15 \\
(\emph{B}) & 10 & 5 & 15 \\
(\emph{C}) & 8 & 7 & 15 \\
(\emph{D}) & 15 & 15 & 30 \\
\hline
\end{tabular}
\caption{\label{tab:design}Configurations studied to compare the performance of an algorithm for automated fetal brain tissue segmentation. For each configuration, the respective number of clinical cases and simulated subjects, as well as the total number of subjects involved in the cross-validation are reported.}
\end{table}

\subsubsection*{Training strategy}
Networks are fed with 64-by-64 overlapping patches in the axial orientation. To avoid any bias, we ensure an equivalent number of 2D patches between an SR-reconstructed clinical case and three orthogonal LR images simulated for a given subject. Intensities of all image patches are standardized and each patch is repeated once using conventional data augmentation operations, namely by random flip and rotation of the patches ($n$ times, by $n\times90$\degre{}, $n\in\llbracket0,3\rrbracket$).

Five-fold cross-validation approaches (training sets: 12 subjects, validation sets: 3 subjects) are used to determine the epochs for the learning rate decay in each configuration.

All networks are trained with an initial learning rate of $0.001$ and a decay factor of $0.1$. (\emph{A}) networks, respectively (\emph{B}), (\emph{C}), and (\emph{D}) networks, are trained for $115$ epochs, respectively $138$, $158$, and $190$ epochs, with a learning rate decay scheduled at epochs $[55,75]$, respectively $[77,109]$, $[95,130]$, and $[94,160]$.

\subsubsection*{Analysis}
The performance of the fetal brain tissue segmentation networks is evaluated based on the Dice similarity coefficient (DSC)\cite{dice_measures_1945} that quantifies the overlap between the predicted segmentation and the manually-annotated ground truth. The performance metrics are assessed on the validation sets. We report the average over all folds. \emph{P}-values of Wilcoxon rank sum test between each configuration and the baseline (\emph{A}) for individual fetal brain tissue segmentation are adjusted for multiple comparisons using Bonferroni correction. \emph{p} $< 0.05$ is considered statistically significant.

\section*{Results}

\subsection*{Computational performance}

As highlighted in Figure \ref{fig:simulation_pipeline} for a fetus of 30 weeks of GA whose brain is covered by twenty-five slices, the computation time to convert segmented HR anatomical images of the fetal brain to MR contrast is in the order of one second. EPG simulations\cite{weigel_extended_2015,weigel_epg_2021} are run in every voxel of the 3D HR anatomical images in less than four minutes. For one axial series with either little, moderate or strong motion, k-space sampling takes less than seven minutes for HASTE images simulated with an acceleration factor of two, respectively less than eight minutes for SS-FSE images simulated without using any acceleration technique.

\subsection*{Qualitative assessment}

\begin{figure}[hbt!]
\centering
\includegraphics[width=0.7\linewidth]{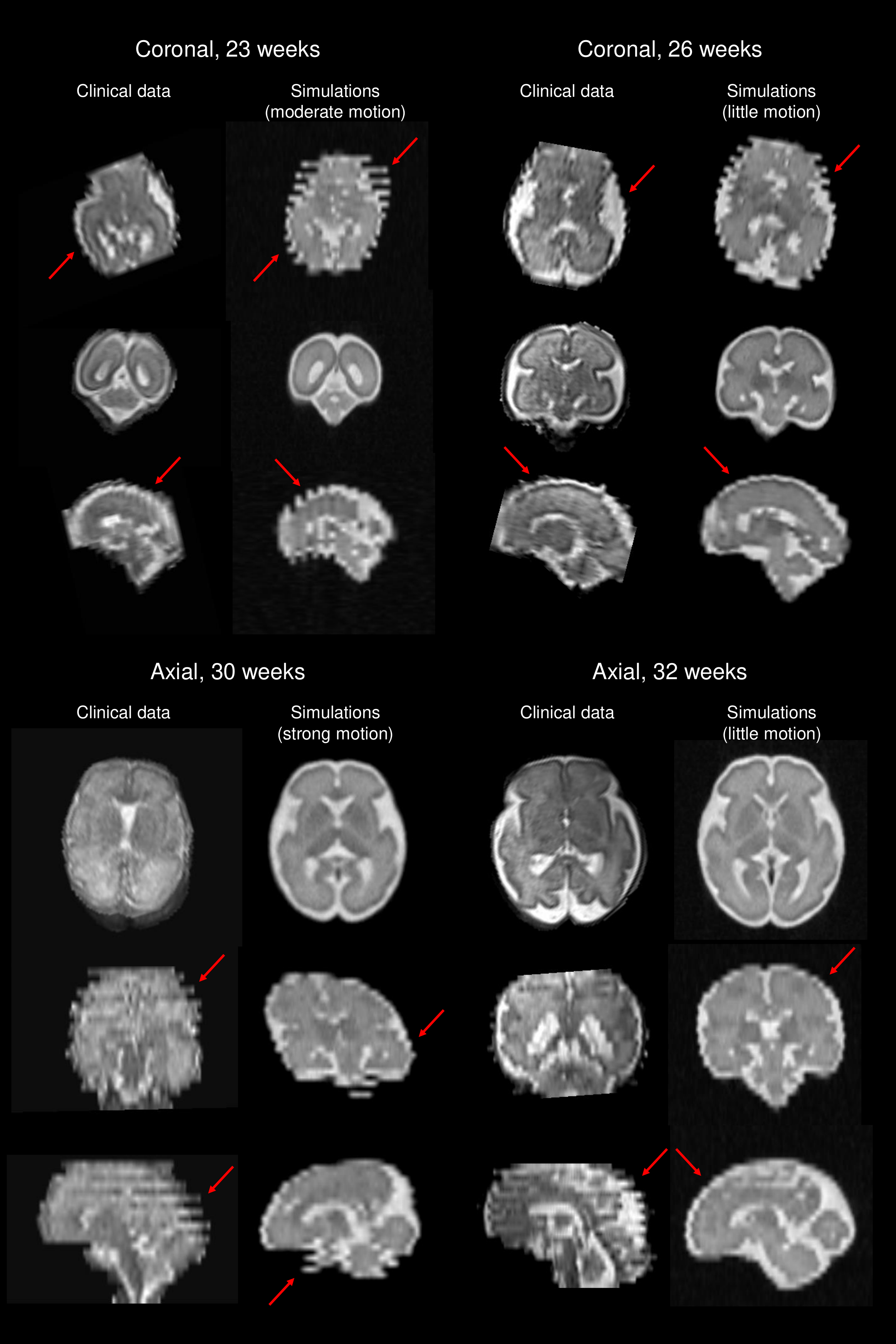}
\caption{Visual inspection and comparison between clinical MR acquisitions and representative simulated HASTE images of the fetal brain in the three orthogonal orientations at four different GA (23, 26, 30 and 32 weeks). The amplitude of movement of the fetus is indicated from the motion index computation. Red arrows point out typical out-of-plane motion patterns.}
\label{fig:visual_inspection}
\end{figure}

Figure \ref{fig:visual_inspection} illustrates the close resemblance between simulated HASTE images of the fetal brain and clinical MR acquisitions in terms of MR contrast between tissues, SNR, brain anatomy and relative proportions across development for representative subjects in the GA range of 23 to 32 weeks, as well as typical out-of-plane motion patterns related to the interleaved slice acquisition scheme.
Experts in neuroradiology and in pediatric (neuro)radiology report a good contrast between gray and white matter, which is important to investigate the cortex continuity and identify the deep gray nuclei as well as any neuronal migration defect. They also notice good SNR in the different series and report proper visualization of the main anatomical structures: the four ventricles, the corpus callosum, the vermis, the cerebellum, even sometimes the fornix. Besides, they are able to monitor the evolution of normal gyration throughout gestation.
However, they point out that small structures such as the hypophyse, the chiasma, the recesses of the third ventricle, and the vermis folds that look part of the cerebellum, are more difficult to observe. The cortical ribbon is clearly visible but quite pixelated, which is likely to complicate the diagnosis of polymicrogyria. White matter appears too homogeneous, which makes its multilayer aspect barely distinguishable, with an MR signal that is constant across GA, thus preventing physicians from exploring the myelination process throughout brain maturation.

\subsection*{Application 1: SR reconstruction}

\subsubsection*{Regularization setting}
\begin{figure}[hbt!]
\centering
\includegraphics[width=0.8\linewidth]{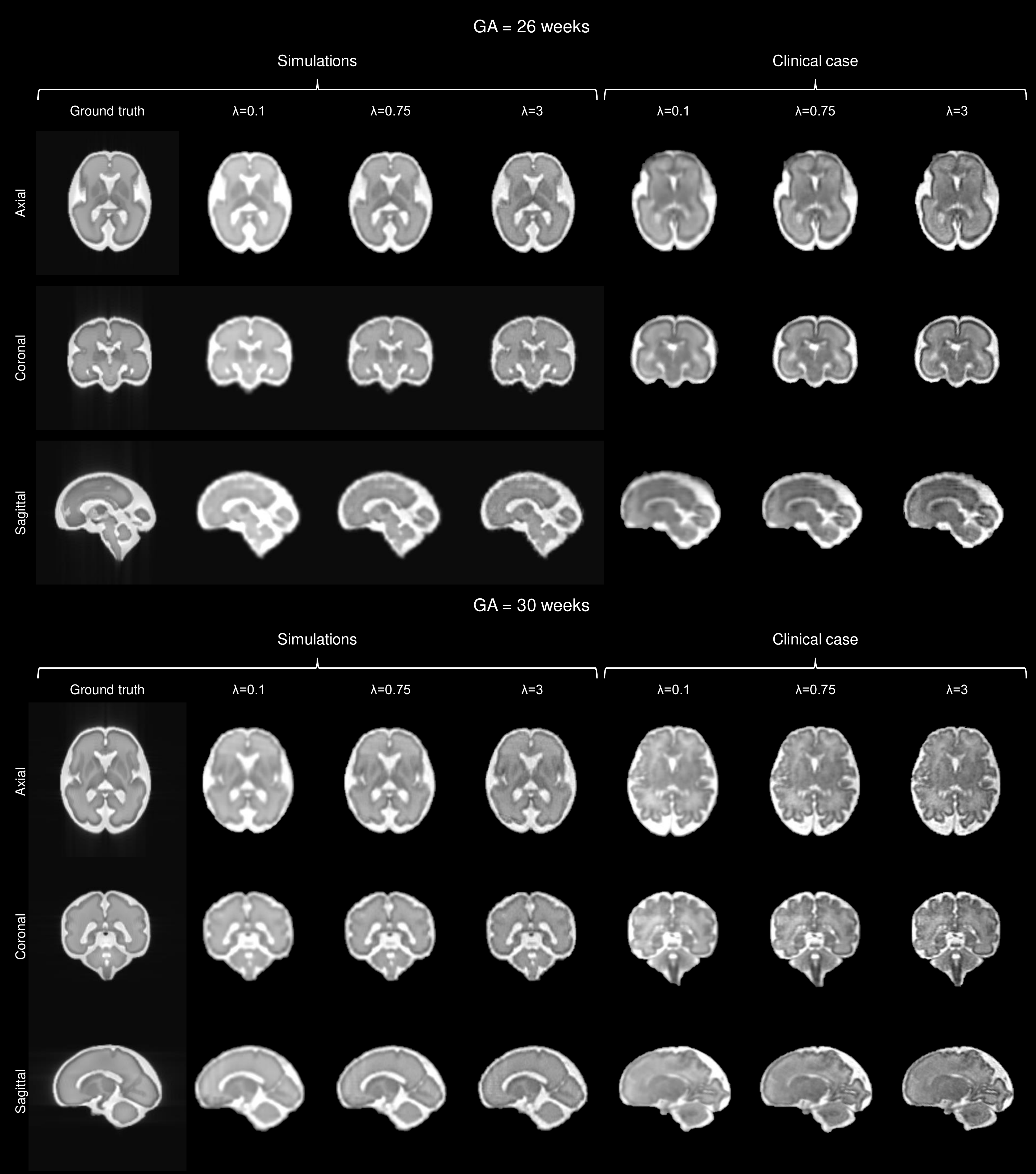}
\caption{Appreciation of the quality of SR reconstruction depending on the weight $\lambda$ that controls the strength of the TV regularization. The potential of our framework FaBiAN for optimizing the reconstruction quality through parameter fine-tuning in the presence of motion is illustrated at two GA: 26 and 30 weeks. Two representative clinical cases are provided for comparison. The results for three values of $\lambda$ are presented. For $\lambda=0.1$, the SR reconstruction looks blurry with poor tissue contrast. Using $\lambda=3$ improves the contrast but the images look noisy. For $\lambda=0.75$, the SR reconstruction is sharp with a contrast between different brain tissues similar to that observed in the 3D isotropic ground truth. Clinical cases from which the simulated HASTE images are derived highlight the accuracy of a SR reconstruction for this intermediate value of $\lambda$, especially with regards to the definition of the corpus callosum and the delineation of the cortex.}
\label{fig:SR_refinement_qualitative_assessment}
\end{figure}

\begin{figure}[hbt!]
\centering
\includegraphics[width=0.7\linewidth]{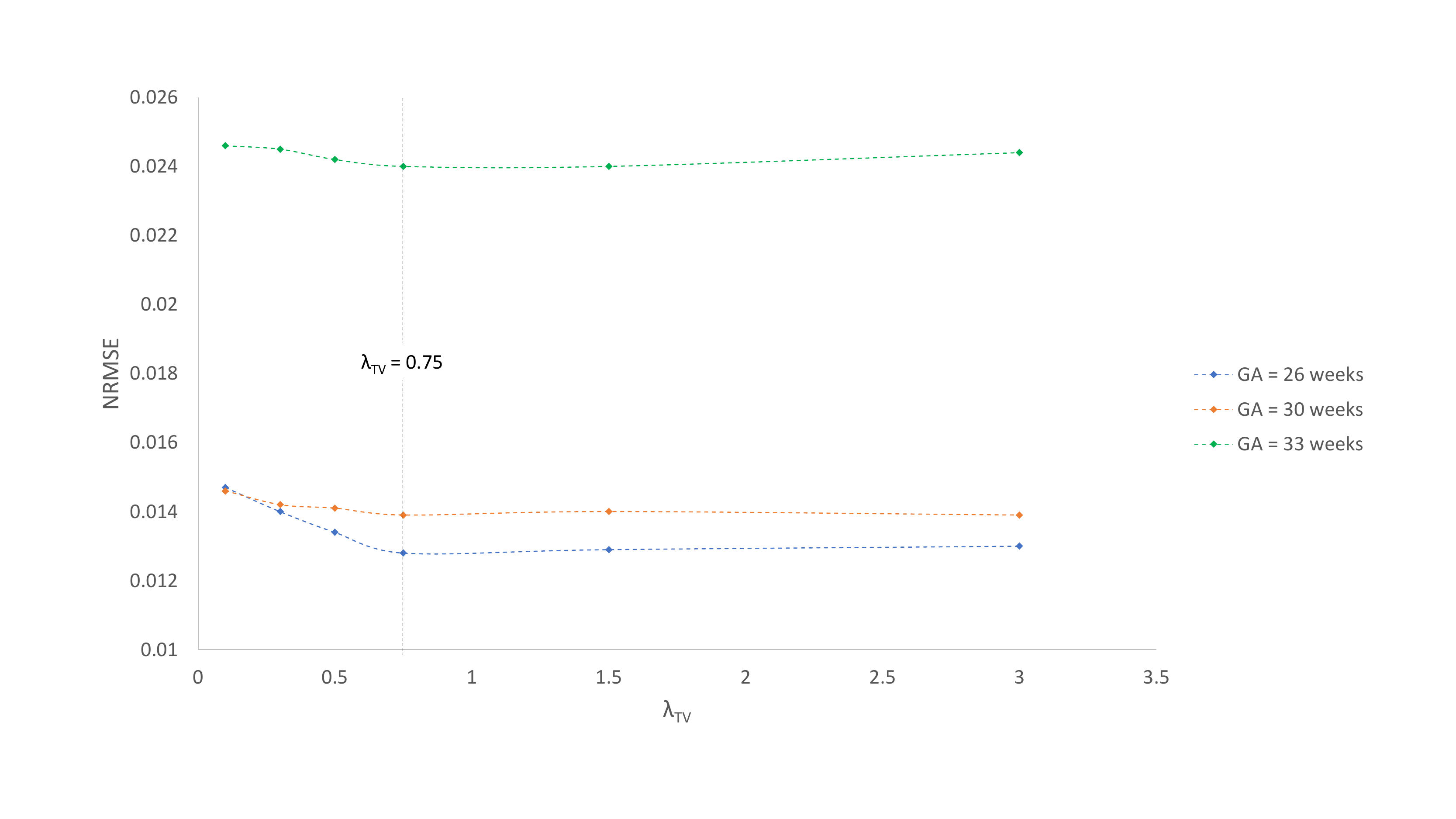}
\caption{NRMSE between SR reconstructions from simulated data at a GA of 26, 30 and 33 weeks respectively and the corresponding 3D HR ground truth depending on the weight $\lambda$ of the TV regularization. Six values of $\lambda$ are tested: 0.1, 0.3, 0.5, 0.75, 1.5 and 3. The NRMSE is minimal for $\lambda$ = 0.75.}
\label{fig:SR_refinement_NRMSE_curves}
\end{figure}

Thanks to its controlled environment, FaBiAN makes it possible to adjust the parameter $\lambda$ for optimal SR reconstruction with respect to a simulated 3D isotropic HR ground truth of the fetal brain.
Figure \ref{fig:SR_refinement_qualitative_assessment} explores the quality of SR fetal brain MRI from LR HASTE images corrupted by motion depending on the weight of TV regularization in two subjects of 26 and 30 weeks of GA respectively. Based on the simulations, a high level of regularization ($\lambda=0.1$) provides a blurry SR reconstruction with poor contrast between the various structures of the fetal brain, especially in the deep gray nuclei and the cortical plate. In addition, the cerebrospinal fluid appears brighter than in the reference image. A low level of regularization ($\lambda=3$) leads to a better tissue contrast but increases the overall amount of noise in the resulting SR reconstruction. A fine-tuned regularization ($\lambda=0.75$) provides a sharp reconstruction of the fetal brain with a high SNR and a tissue contrast close to the one displayed in the reference image.
In the SR images reconstructed from clinical LR HASTE series altered by a little-to-moderate level of motion, as in the simulations, the structure of the corpus callosum and the delineation of the cortex are especially well defined for appropriate TV regularization ($\lambda=0.75$), leading to high-SNR HR images of the fetal brain.
Although the NRMSE between SR reconstructions from simulated HASTE images and the corresponding ground truth are close to each other for a given GA across the various weights studied, Figure \ref{fig:SR_refinement_NRMSE_curves} shows that the error is systematically minimal for $\lambda=0.75$, which further supports this parameter setting for optimal SR reconstruction of the fetal brain from this type of MR images.

\subsubsection*{Number of LR series: an SNR and motion case study}
\begin{figure}[hbt!]
\centering
\includegraphics[width=\textwidth]{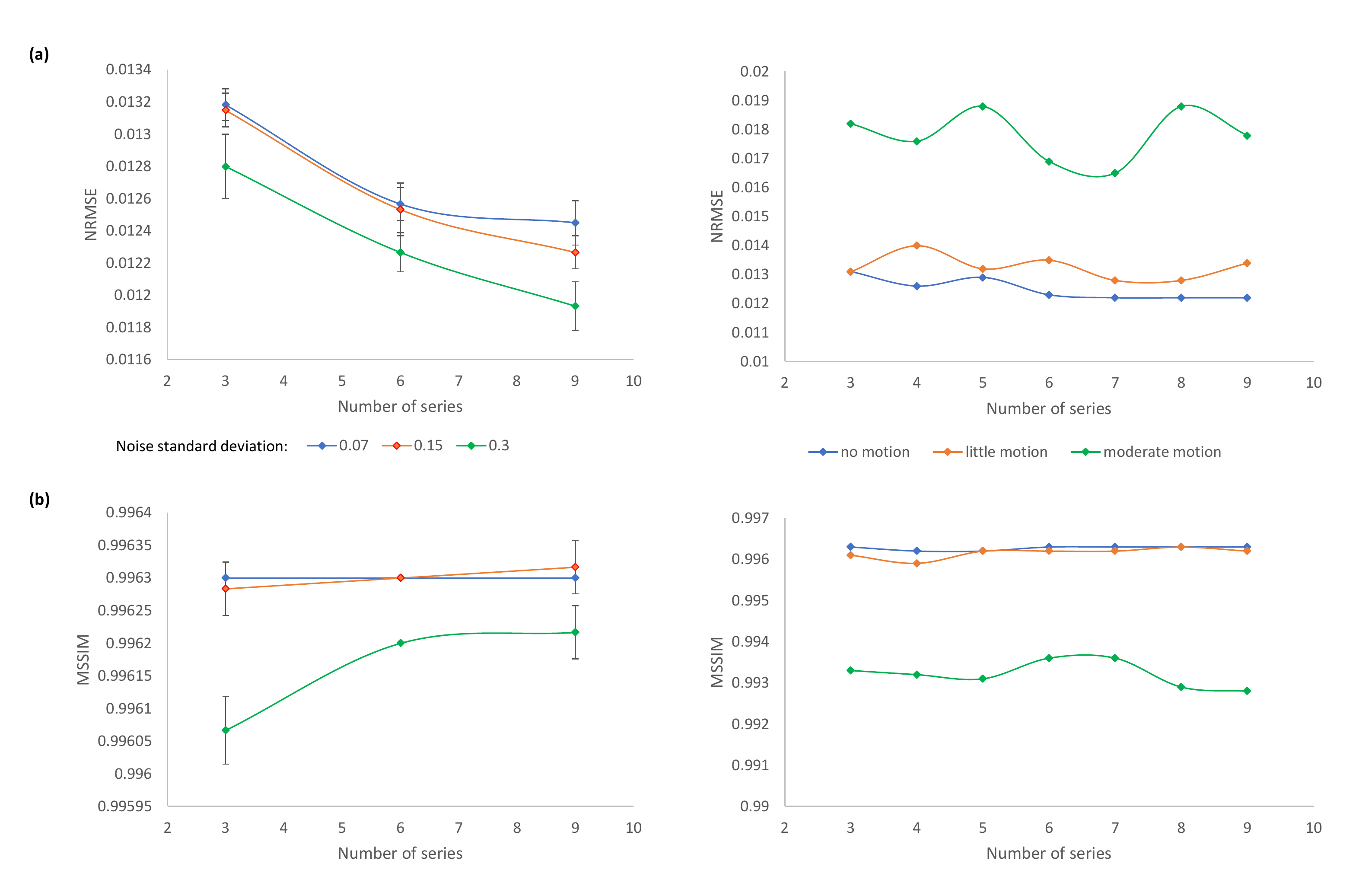}
\caption{(a) NRMSE and (b) MSSIM between SR reconstructions from different numbers of orthogonal LR HASTE series simulated at a GA of 30 weeks and the corresponding static 3D HR ground truth. The left panel shows results for motion-free data with various noise levels, a standard deviation of 0.15 leading to a similar appearance as in clinical acquisitions. The right panel illustrates how the algorithm performs depending on the amplitude of fetal movements in the input series.}
\label{fig:SR_nb_series_plots}
\end{figure}

Figure \ref{fig:SR_nb_series_plots} shows the NRMSE and the MSSIM between SR reconstructions from different numbers of orthogonal LR HASTE series simulated with various SNR or variable amplitude of movements and a 3D HR reference\cite{lajous_magnetic_2021}.

In the case of static data (Figure \ref{fig:SR_nb_series_plots}-left panel), the NRMSE decreases when increasing the number of series used in the SR reconstruction. According to both the NRMSE and the MSSIM, the quality of the SR reconstructions resulting from simulated images with an SNR close to that observed in clinical acquisitions and from synthetic images with a distribution of complex Gaussian noise of twice less standard deviation is similar. Noisier images lead to a slight decrease in the NRMSE, but also in the MSSIM which in turn increases with the number of series.

The stronger the movements in the LR series, the higher the NRMSE of the resulting SR reconstruction. The addition of motion-corrupted LR series to reconstruct a SR volume of the fetal brain does not increase the MSSIM. Data with slight motion are well handled by the SR algorithm as the MSSIM is equivalent for SR reconstructions from static images and series with little motion. In the case of moderate motion, the MSSIM is lower than in the case of little motion.

\begin{figure}[hbt!]
\centering
\includegraphics[width=0.6\linewidth]{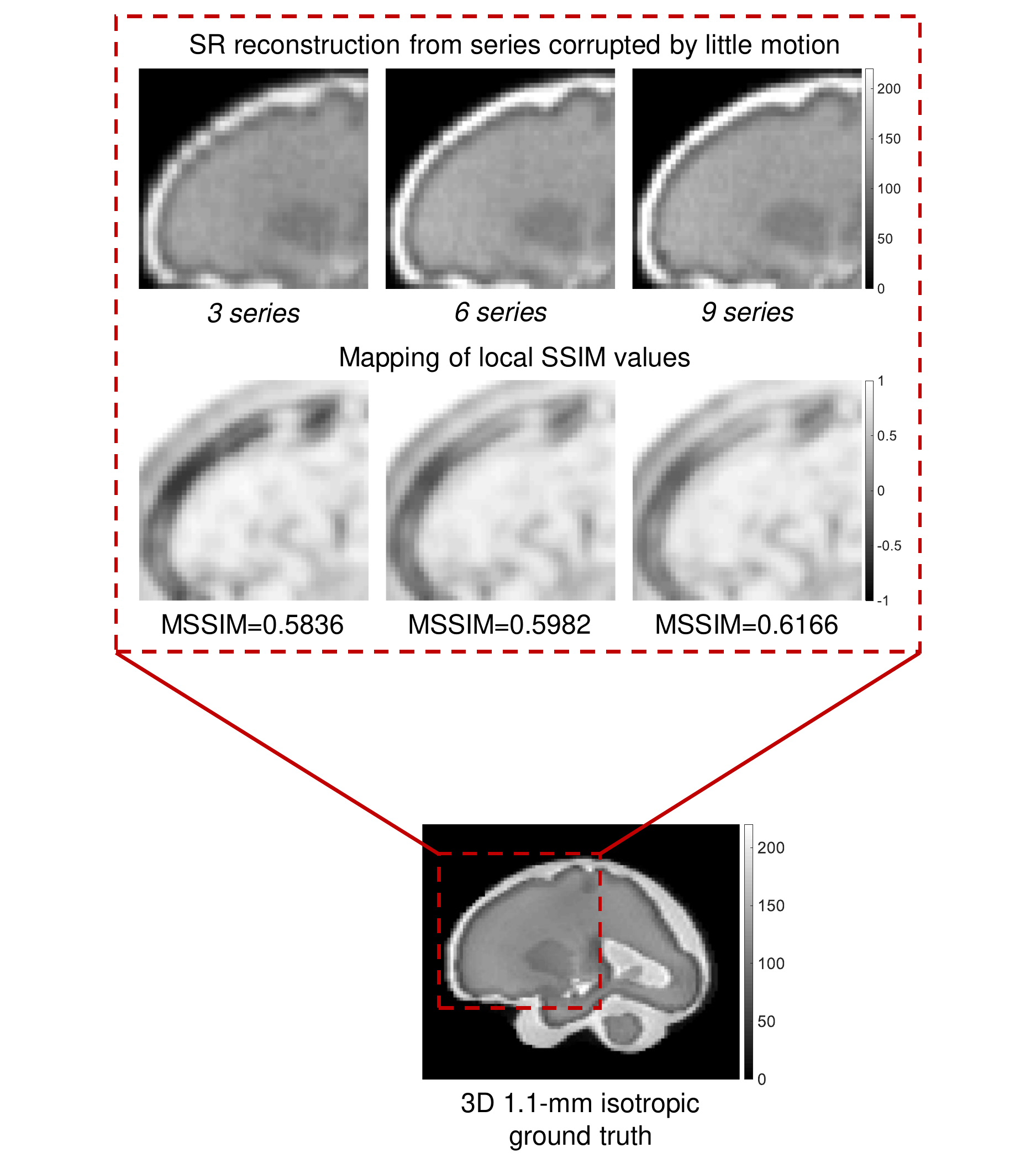}
\caption{Appreciation of sharpness and tissue contrast enhancement in SR reconstructions from higher numbers of simulated orthogonal LR HASTE images corrupted by little motion at a GA of 30 weeks in comparison with the corresponding static 3D HR ground truth. The frontal cortex looks smoother and the putamen area sharper in the SR reconstruction from nine series compared to the SR reconstruction from three series. The mapping of local SSIM values and the computation of the MSSIM over the corresponding region-of-interest further support these observations.}
\label{fig:SR_nb_series_visual}
\end{figure}

Figure \ref{fig:SR_nb_series_visual} highlights the benefit of increasing the number of orthogonal LR series on the rendering of the SR reconstruction. The higher the number of LR series combined in the SR reconstruction, even altered by motion, the smoother the frontal cortex and the sharper the putamen area in the resulting SR volume\cite{lajous_magnetic_2021}.

\subsection*{Application 2: Data augmentation for automated fetal brain tissue segmentation}

We aim at proving the realistic appearance of the simulated images (representative SS-FSE images of two subjects scanned / simulated at 1.5 T and 3 T respectively are provided as Supplementary Material) and demonstrating the practical value of the developed environment to complement clinical datasets for data augmentation strategies in deep learning, here with the example of fetal brain tissue segmentation.

Table \ref{tab:DSC_tissue_overall} shows the mean DSC $\pm$ standard deviation computed for every segmented brain tissue in each configuration. Overall, the performance of the segmentation algorithm is maintained when replacing original subjects by synthetic images obtained from an SS-FSE sequence simulated with the same acquisition parameters as in the clinical protocol. This trend is also observed for each individual structure studied. In the configuration (\emph{C}), the overall DSC for brain tissue segmentation is slightly increased as compared to the baseline with statistical significance.

\begin{table}[hbt!]
\centering
\begin{tabular}{r l l l l}
\hline
 & (\emph{A}) C15/S0 & (\emph{B}) C10/S5 & (\emph{C}) C8/S7 & (\emph{D}) C15/S15 \\
\hline
CSF \& Ventricles & 0.93 $\pm$ 0.01 & 0.94 $\pm$ 0.01 & 0.94 $\pm$ 0.02 (*) & \textbf{0.95} $\pm$ 0.02 (*) \\
Cortical GM & 0.77 $\pm$ 0.02 & 0.80 $\pm$ 0.04 & 0.81 $\pm$ 0.05 (*) & \textbf{0.84} $\pm$ 0.05 (*) \\
WM & 0.92 $\pm$ 0.01 & 0.92 $\pm$ 0.02 & 0.92 $\pm$ 0.02 & \textbf{0.93} $\pm$ 0.02 \\
Cerebellum & 0.88 $\pm$ 0.04 & 0.87 $\pm$ 0.10 & 0.87 $\pm$ 0.09 & \textbf{0.92} $\pm$ 0.04 (*) \\
Deep GM & 0.85 $\pm$ 0.03 & 0.84 $\pm$ 0.10 & 0.87 $\pm$ 0.04 & \textbf{0.90} $\pm$ 0.04 (*) \\
Brain stem & 0.84 $\pm$ 0.03 & 0.85 $\pm$ 0.04 & 0.86 $\pm$ 0.04 & \textbf{0.88} $\pm$ 0.03 (*) \\
Overall & 0.87 $\pm$ 0.06 & 0.87 $\pm$ 0.08 & 0.88 $\pm$ 0.06 (*) & \textbf{0.90} $\pm$ 0.05 (*) \\
\hline
\end{tabular}
\caption{\label{tab:DSC_tissue_overall} DSC (mean $\pm$ standard deviation) in the different configurations studied for all segmented brain tissues: cerebrospinal fluid (CSF) and ventricles, cortical gray matter (GM), white matter (WM), cerebellum, deep gray matter and brain stem, and on average. The number of clinical cases (Cxx) and the number of simulated subjects (Sxx) are recalled. The segmentation algorithm performs better (score in bold) in every structure when complementing the baseline dataset (configuration \emph{A}) with simulated subjects (configuration \emph{D}).
\emph{P}-values of Wilcoxon rank sum test between each configuration and the baseline (\emph{A}) for individual fetal brain tissue segmentation are adjusted for multiple comparisons using Bonferroni correction. \emph{p} $< 0.05$ (*) is considered statistically significant.}
\end{table}

Since the simulated images look realistic enough to substitute for original clinical acquisitions, we further investigate if they can be used to complement real data in a data augmentation strategy. As expected, increasing the training data by a combination of fifteen original cases and fifteen simulated subjects (configuration \emph{D}) results in a significantly improved mean DSC of 0.90 $\pm$ 0.05 over the six segmented brain tissues as compared to the baseline (\emph{A}) (see Table \ref{tab:DSC_tissue_overall}). In more detail, the DSC is higher for all segmented brain structures when complementing clinical acquisitions with close simulated data, with statistical significance ($\emph{p} < 0.05$) for the cerebrospinal fluid and ventricles, the cortical gray matter, the cerebellum, the deep gray matter and the brain stem.

\begin{figure}[hbt!]
\centering
\includegraphics[width=\textwidth]{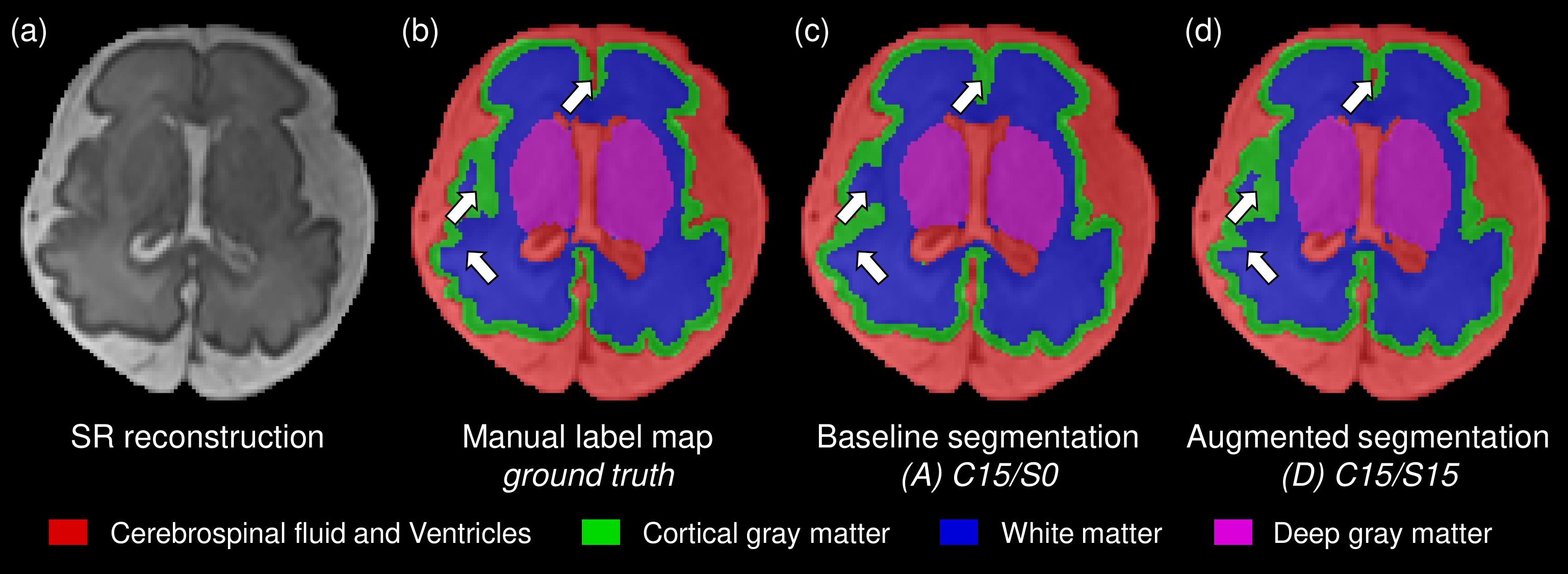}
\caption{Illustration of the accuracy of fetal brain tissue segmentation in a subject of 30.6 weeks of GA on (a) an axial slice from the SR reconstruction. Comparison of (b) the reference manual annotations, (c) the segmentation results obtained by the baseline (\emph{A}) where the network is trained on clinical SR reconstructions only (C15/S0), (d) the segmentation results obtained by the configuration (\emph{D}) that complements this original dataset with fifteen additional simulated subjects (C15/S15), overlaid on the SR image. The segmentation of the cortex especially looks more accurate in (d), with an increased sensitivity to folding as highlighted by the white arrows.}
\label{fig:segmentation}
\end{figure}

Figure \ref{fig:segmentation} illustrates on an axial view the accuracy of fetal brain tissue segmentation in a subject of 30.6 weeks of GA. The results obtained in the configuration (\emph{D}) where real clinical data are complemented with simulated subjects look close to the manually-annotated ground truth. In particular, the segmentation seems to be more accurate in the cortex with an enhanced sensitivity to folding as compared to the segmentation obtained in the baseline (\emph{A}) where the network is solely trained on clinical data.

\section*{Discussion \& Conclusion}

In this work, we present FaBiAN, a novel Fetal Brain magnetic resonance Acquisition Numerical phantom, and illustrate some of its potential uses. Our tool relies on EPG simulations\cite{weigel_extended_2015,weigel_epg_2021} to account for stimulated echoes in the computation of the T2 decay in every voxel of the HR anatomical images from which the simulated images are derived. The developed framework remains general and highly flexible in the choice of the sequence parameters and anatomical settings available to the user. It simulates as closely as possible the physical principles involved in FSE sequences on several MR systems, namely from various MR vendors and at different magnetic field strengths, resulting in highly realistic T2w images of the developing brain throughout gestation.

The limitations in the resemblance of the simulated HASTE images as compared to typical clinical MR acquisitions may be explained by the origin of the simulated images and the lack of T1 and T2 ground truth measurements, both in the multiple fetal brain tissues and throughout maturation. HASTE images are simulated from a normative spatiotemporal MRI atlas of the fetal brain\cite{gholipour_normative_2017} where representative images at each GA correspond to an average of fetal brain scans across several subjects, thus resulting in smoothing of subtle inter-individual heterogeneities, especially in the multilayer aspect of the white matter. As a first approximation, we consider average T1 and T2 values of the various fetal brain structures labeled as gray matter, white matter or cerebrospinal fluid throughout development. As a result, our simulated images may fail to capture the fine details of the fetal brain anatomy as well as maturation processes that imply changes in T1 and T2 relaxation times during gestation. In this sense, experts report that they would feel confident in performing standard biometric measurements on the simulated images and in evaluating the volume of white matter, but not its fine structure.
It is worth noticing that in-plane motion artefacts like signal drops are not accounted for at this stage, as slices severely corrupted by such artefacts will be removed from the analysis.
Furthermore, we restricted our simulations to images of the fetal brain without the remaining fetal anatomy or the surrounding maternal uterine cavity, as most image analysis and post-processing techniques intended for fetal brain MRI, including SR reconstruction, first perform an automated fetal brain extraction/segmentation. Unlike other models that represent fetal subjects by scaled pediatric data and oversimplify the complexity of the fetal brain anatomy by reducing it to white matter alone or to a single brain tissue label\cite{christ_virtual_2010,gosselin_development_2014,hasgall_itis_2018}, we wanted our numerical simulations to be based on a comprehensive model of the fetal brain consistent with the underlying clinical application. In this sense, the normative spatiotemporal MRI atlas of the fetal brain\cite{gholipour_normative_2017} from which the simulated FSE images are built captures the details of the fetal brain anatomy throughout maturation. Thanks to the flexibility of FaBiAN, the surrounding fetal and maternal anatomy, as well as any other structures, could easily be included as long as we have access to segmented HR images, for instance by resorting to more complex anatomical models such as the XCAT phantoms\cite{segars_mcat_2010,segars_4d_2010,norris_set_2014}. Several models can also be combined.

We have designed this open-source simulator to aid in the development and validation of advanced image processing techniques dedicated to improving fetal brain MRI.
Despite SR reconstruction has already demonstrated its potential for accurate biometric measurements in the fetal brain\cite{velasco-annis_normative_2015,pier_3d_2016,khawam_fetal_2021}, some parameters still need to be adjusted to the nature of the input LR images to provide optimal evaluation and support computer-assisted diagnosis.
In fetal MRI, the level of regularization is commonly set empirically based on visual perception\cite{gholipour_robust_2010,rousseau_super-resolution_2010,kuklisova-murgasova_reconstruction_2012,ebner_automated_2020}. Intuitively, the level of regularization depends on the amount of data available to solve the ill-posed inverse problem. Thanks to its controlled environment, the presented framework makes it possible to explore the optimal settings for SR fetal brain MRI according to the quality of the input motion-corrupted LR series with respect to a synthetic 3D HR ground truth. It is worth noticing that in-plane motion artefacts like signal drops are not accounted for in the simulation workflow at this stage, as heavily corrupted slices are commonly removed from the reconstruction.
Besides, FaBiAN also enables quantitative assessment of the robustness of any SR reconstruction algorithm depending on various parameters that can be intrinsic to the system like noise, or related to the clinical application such as the amplitude of fetal motion in the womb and the number of series used for SR reconstruction\cite{lajous_magnetic_2021}. For instance, a decrease in the MSSIM when adding motion-corrupted series to reconstruct a SR volume of the fetal brain as compared to the SR reconstruction from static series can be caused by an inappropriate slice-to-volume registration (SVR). Therefore, our numerical phantom provides a valuable framework for reproducibility studies and validation of image processing methods.
Additional examples among its wide variety of applications include the simulation of a static reference volume at a given GA on which to align the clinical orthogonal LR series acquired in a subject at the same GA in order to perform SVR and subsequent SR reconstruction of the fetal brain, especially in the presence of heavily motion-corrupted acquisitions. Besides, the performance of the SVR can be quantitatively assessed by comparing the motion transform estimated by the algorithm to the controlled 3D rigid movements actually simulated in the images. Synthetic HR images can also be used as a reference for more general motion compensation techniques.

As raised by Wissmann and colleagues, the lack of comparability between simulation setups hinders the evaluation of image reconstruction methods in relation to each other\cite{wissmann_mrxcat_2014}. With this first MR acquisition simulation framework of the developing fetal brain, we aim at providing the community with a unified environment for the evaluation and validation of various post-processing techniques to improve the analysis of fetal brain MR images and support accurate diagnosis.

Furthermore, the developed simulator generates T2w images of the fetal brain realistic enough to complement real clinical acquisitions for data augmentation strategies, as shown here with a proof of concept for fetal brain tissue segmentation. It especially makes it possible to exploit the whole range of GA in the simulated data when clinical cases are scarce and not necessarily uniformly distributed across development. Thus, we can take advantage of larger and more diverse datasets at no cost.

In the configuration (\emph{C}) that combines eight original cases and seven simulated subjects, the mean DSC over all segmented fetal brain tissues is slightly increased as compared to the baseline. This can be explained by the higher number of synthetic images contained in the validation sets.
Although the resort to an atlas restricts the inter-subject variability at a given GA, the great flexibility of FaBiAN also lies in the possibility of simulating images from various sources, either atlases or clinical segmented HR anatomical images of the fetal brain like SR reconstructions. Thus, it becomes possible to simulate several subjects at a given GA from various fetal brain models to increase the inter-subject anatomical heterogeneity in the synthetic images.

In line with the demonstration of the added value of our numerical framework FaBiAN for diverse applications that revolve around improving diagnosis and prognosis from MR images of the developing fetal brain, future work aims at investigating the ability of such a simulator to generalize post-processing tools like fetal brain tissue segmentation to datasets acquired on other MR systems and with other sequence parameters using a collection of various synthetic images for domain adaptation techniques.
\\


\textbf{Data availability -} All the simulated datasets that support the various proof-of-concept studies documented in this manuscript will be made publicly available on a Zenodo repository upon acceptance of the present work.

\bibliography{HASTE_simulator}


\section*{Acknowledgements}

This work was supported by the Swiss National Science Foundation through grants 182602, 141283 and 173129, by the European Union’s Horizon 2020 research and innovation program under the Marie Sklodowska-Curie project TRABIT (agreement No 765148), by the Forschungszentrum für das Kind (FZK), the Hasler Foundation, the Anna Muller-Grocholski Foundation, and the Zurich Neuroscience Center PhD Grant. We acknowledge access to the facilities and expertise of the CIBM Center for Biomedical Imaging, a Swiss research center of excellence founded and supported by Lausanne University Hospital (CHUV), University of Lausanne (UNIL), Ecole polytechnique fédérale de Lausanne (EPFL), University of Geneva (UNIGE) and Geneva University Hospitals (HUG).
We would like to thank Dr. Ruth Tuura from the Center for MR Research of University Children’s Hospital Zurich (Zurich, Switzerland) for her support in answering questions about acquisition parameters specific to the implementation of the SS-FSE sequence by GE. We would also like to thank Ji Hui from the Center for MR Research of University Children’s Hospital Zurich (Zurich, Switzerland) for the annotation of the SR fetal brain MR images in subjects scanned at Kispi.

\section*{Author contributions statement}
H.L. investigated the HASTE sequence pulse design using the IDEA sequence programming interface provided by Siemens Healthineers, built on the initial simulation framework to develop FaBiAN, implemented additional features such as the inclusion of intensity non-uniformity fields and stochastic motion, and extended this numerical phantom, originally thought to simulate HASTE acquisitions of the fetal brain, to FSE sequences from other MR vendors. H.L. suggested a classification of segmented brain tissues as gray matter, white matter or cerebrospinal fluid, reviewed the state of the art in order to find out T1, T2 and T2* values reported in the developing fetal brain as well as in premature and full-term newborns at 1.5 T, and extrapolated from the more general literature the corresponding T1 and T2 values at 3 T. H.L. automatically generated the brain masks of the simulated images and manually refined them, reconstructed the super-resolution fetal brain volumes from clinical, respectively synthetic, 2D motion-corrupted series, participated in the design of the different case studies, analyzed all the results, prepared the tables and figures, and wrote the article. C.W.R. implemented the skeleton of the original simulator, helped H.L in optimizing it, and suggested complementary case studies. T.H. helped H.L. to understand the design of the HASTE sequence and to improve the simulation process. C.W.R. and T.H. equally contributed to this work. P.d.D. provided clinical cases acquired at CHUV from the fetal MRI database whose SR reconstructions were previously evaluated as excellent. P.d.D and H.K. participated in the design of the data augmentation experiments using simulated images in the context of fetal brain tissue segmentation. P.d.D. implemented the deep learning network for fetal brain tissue segmentation and evaluated its performance in the different configurations studied. P.d.D. and S.T. assisted H.L. in using the SR reconstruction pipeline\cite{tourbier_medical-image-analysis-laboratorymialsuperresolutiontoolkit_2020}. S.T. especially helped with the installation procedure. Y.A.G. provided a script that saves matlab matrices to nifti images with the right orientation matrix. J.Y. provided a script that preprocesses the data in k-space using a Fermi low-pass filter before zero-interpolation filling. T.Y. provided a script that computes a motion index from the brain masks. K.P. shared the data from fifteen subjects scanned at Kispi, including clinical LR SS-FSE images of the fetal brain and corresponding SR reconstructions together with their manual annotations. K.P. performed the SR reconstructions of the data from Kispi. K.P. also provided information about the protocol for fetal brain examination as it is conducted in clinical routine at Kispi. J.B.L. provided information on the clinical protocol for fetal brain MRI as it is performed at CHUV. R.M. assessed the clinical cases. A.J. supported the use of detailed clinical cases from Kispi to extend our simulation framework to FSE sequences from other MR vendors and demonstrate its added value in providing multiple synthetic data at no cost for data augmentation strategies. M.K. and V.D. validated and complemented the labels of the various anatomical structures in the developing fetal brain. V.D. provided approximate relaxometric T1 and T2 values in the cerebrospinal fluid in pediatric patients. M.K. detailed the features radiologists routinely analyze in clinical images of the fetal brain to determine the neurotypical or pathological status of the fetus, as well as the various artefacts they typically encounter in LR HASTE images. M.K. and V.D. provided qualitative assessment of the simulated HASTE images. P.H., T.K. and M.S. supported the writing of this manuscript and provided regular feedback during the project. M.B.C. regularly assessed the realistic appearance of the simulated HASTE images as compared to typical clinical acquisitions of the fetal brain at CHUV throughout the implementation of the simulation pipeline, qualitatively estimated the level of motion in clinical series, suggested some application examples where the simulator could be particularly valuable, participated in the design of the various experiments presented in this manuscript, and supported the analysis of the results as well as the writing of the manuscript. M.B.C. is the recipient of the FNS grant that supports the whole project. All authors reviewed the manuscript.

\section*{Additional information}


\textbf{Competing interests}
The authors declare no competing interests.

\newpage
\section*{Supplementary Material}

\begin{figure}[hbt!]
\centering
\includegraphics[width=0.8\linewidth]{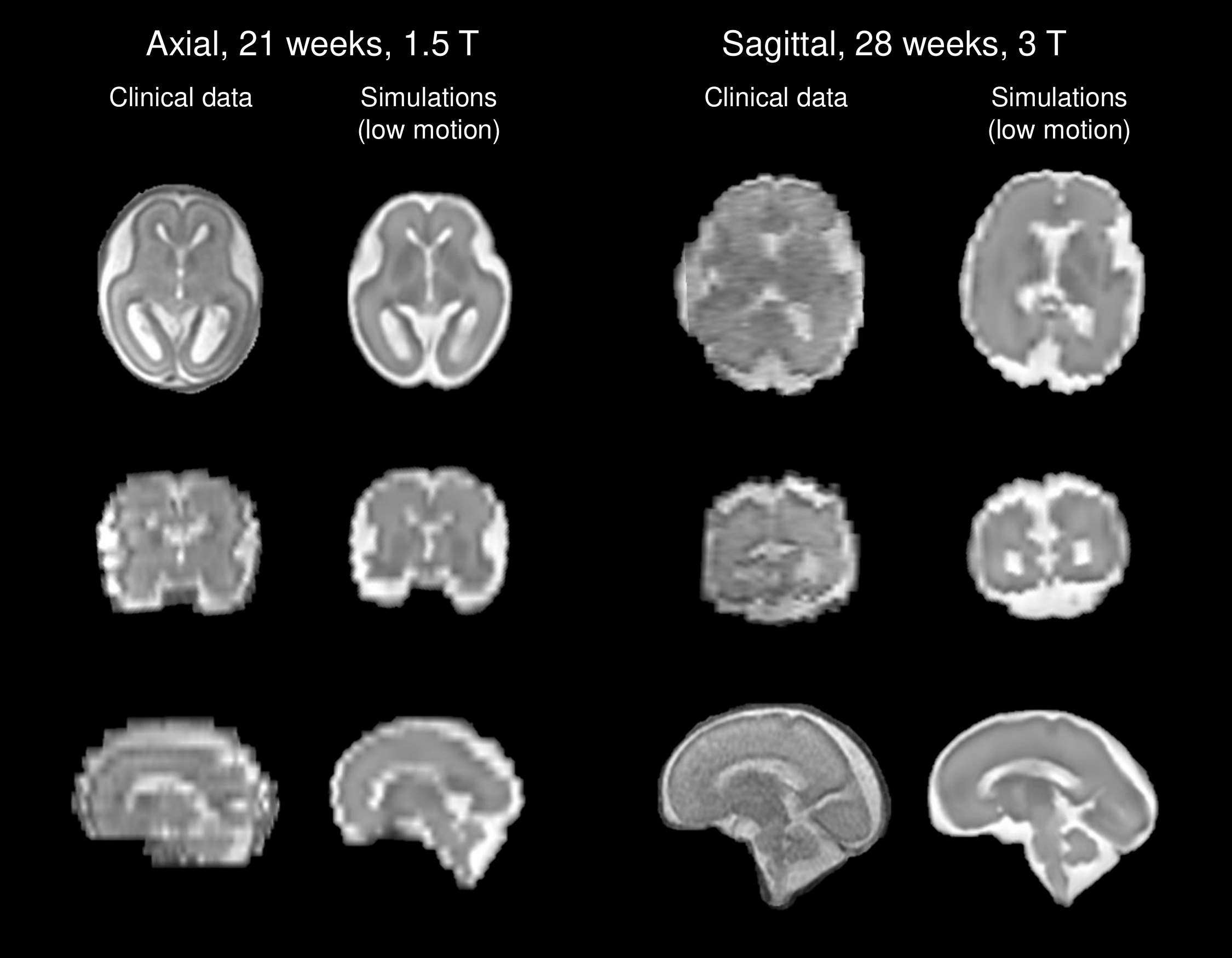}
\caption{Visual inspection and comparison between clinical MR acquisitions (GE Healthcare) and representative simulated SS-FSE images of the fetal brain in the three orthogonal orientations, at 1.5 T and 3 T respectively, and at two GA (21 and 28 weeks).}
\label{fig:SS-FSE_images}
\end{figure}


\end{document}